\documentclass[aps,prl,twocolumn,showpacs,preprintnumbers,amsmath,amssymb]{revtex4}
\usepackage[english]{babel}
\newcommand{\beq}{\begin{equation}}
\newcommand{\eeq}{\end{equation}}
\newcommand{\beqa}{\begin{eqnarray}}
\newcommand{\eeqa}{\end{eqnarray}}
\newcommand{\m}{{\mathbf m}}
\newcommand{\x}{{\mathbf x}}

\newcommand{\h}{{\mathbf h}}
\newcommand{\1}{{\mathbf 1}}
\newcommand{\J}{{\mathbf J}}

\newcommand{\T}{{\mathbf T}}
\newcommand{\M}{{\mathbf M}}

\newcommand{\tr}{\mbox{Tr}}
\renewcommand{\(}{\left(}
\renewcommand{\)}{\right)}
\newcommand{\E}[1]{\left\langle #1 \right\rangle}
\newcommand{\heading}[1]{\vspace{0.5em}{\centering\em #1.}\\}

\begin{document}

\preprint{LU TP 03-12}

\title{Random Graphs with Hidden Color}

\author{Bo S\"oderberg}
\email{Bo.Soderberg@thep.lu.se}
\affiliation{Complex Systems Division, Dept. of Theoretical Physics, Lund University}

\date{\today}% It is always \today, today, but any date may be explicitly specified

\begin{abstract}
We propose and investigate a unifying class of sparse random graph
models, based on a {\em hidden coloring} of edge-vertex incidences,
extending an existing approach, random graphs with a given degree
distribution, in a way that admits a nontrivial correlation structure
in the resulting graphs.
The approach unifies a number of existing random graph ensembles
within a common general formalism, and allows for the analytic
calculation of observable graph characteristics.
%important asymptotic properties.
%
In particular, generating function techniques are used to derive the
size distribution of connected components (clusters) as well as the
location of the percolation threshold where a giant component appears.
\end{abstract}

%\pacs{Valid PACS appear here}% PACS, the Physics and Astronomy Classification Scheme.

%\keywords{Suggested keywords}%Use showkeys class option if keyword display desired

\maketitle

\heading{Introduction}
There is a growing interest in complex networks, in the physics
community as well as in other sciences, partly due to an increased
availability of data on real-world networks.
This is reflected in a rapidly increasing number of
models of random graphs \cite{ErRe60,BenCan,WatStr98,Sod02,BerLae02}
and dynamical random graphs \cite{AlBa99,AlBa00,Call01,Turo02,DoMeSa01},
with varying degrees of generality.

This multitude of models calls for a unifying formalism, including
more specific models as special cases, while allowing for the
calculation of observable characteristics that can be compared to
those of real networks. Dynamical models are interesting in their own
right, but the dynamics is seldom directly observable in real-world
networks, and we will focus on static ensembles of random graphs,
irrespective of whether they result from a dynamical process or not.

Specifically, we will consider models of simple, undirected graphs
that are {\em sparse} (the edge count grows linearly with the node
count $N$) and {\em truly random} (having no underlying regular
structure).
The {\em classic random graph} in its sparse version is of this type
\cite{ErRe60,Jans00,Boll01}, where each of the $N(N-1)/2$ possible
edges is independently and randomly realized with a fixed probability
$p=c/N$. It has a Poissonian asymptotic degree (connectivity)
distribution with average $c$, and a percolation threshold at
$c=1$. It fails, however, to describe most real-world networks.

Instead we turn to two of the more general approaches, based on
slightly different philosophies.
One, to be referred to as {\em DRG} (Degree-driven random graphs),
amounts to chosing a random member from the set of simple labelled
graphs with a given arbitrary degree distribution
\cite{BenCan,MoRe95,Newm01}.
The other is Inhomogeneous random graphs \cite{Sod02}, {\em IRG},
where the classic model is generalized by randomly coloring vertices
according to a color distribution $\{r_i\}$, and realizing edges
independently with color-dependent probabilities $c_{ij}/N$.
Both yield analytically tractable models displaying well-defined
percolation thresholds and degree distributions, both include a number
of more specific models -- and both have limitations: DRG fails to
produce non-trivial edge correlations, as seen in the factorization of
the combined degree distribution of connected vertex pairs
\cite{Call01}; in IRG, the resulting degree distribution is limited to
a mix of Poissonians \cite{Sod02}.

These approaches are not unrelated: The restriction of DRG to degree
distributions in the form of a Poissonian mix is in fact
asymptotically equivalent to the restriction of IRG to a rank-one $c$
matrix, $c_{ab}=C_aC_b$ (exhibiting DRG's lack of correlations); this
common subset contains the classic model \cite{Sod02}.

\heading{Basic Idea}
By combining the philosophies of DRG and IRG, a more general class of
analytically tractable sparse random graph models can be constructed.
This unifying approach, to be referred to as {\em CDRG} (for Colored
DRG), contains IRG and DRG as particular subsets, and
is defined as a direct extension of DRG by assigning a {\em hidden
color} to each vertex connection (a half-edge, or {\em stub}). As a
result each edge will be associated with a pair of colors, one for
each endpoint.
We then consider a given distribution $\{p_{\m}\}$ of {\em colored
degrees} $\m=(m_1\dots m_K)$, where for each vertex its number $m_k$
of stubs of each color $k$ is accounted for, and allow the edge
distribution to be color-sensitive by specifying also the distribution
of edge color pairs. The resulting ensemble of stub-colored graphs
yields, if the coloring is considered {\em unobservable}, a
well-defined graph ensemble.  The coloring thus can be thought of as a
set of {\em hidden variables}, the purpose of which is to induce
correlations in the resulting graphs.

Below, we will discuss the definition and implementation of CDRG
models, derive the asymptotic cluster size distribution yielding
equations for the percolation threshold, and identify the subsets
corresponding to DRG (trivial) and IRG (less trivial).

%%%%%%%%%%%%%%%% Asymptotic def. %%%%%%%%%%%%%%%%%%
%\newpage
\heading{Asymptotic Model Specification}
A particular asymptotic CDRG model is defined by specifying:
\begin{itemize}
\item a definite color space, say $\{1,2,\dots,K\}$;
\item an asymptotic colored degree distribution (CDD), $p_{\m}$,
 defining the relative frequencies of vertices with different colored
 degrees $\m=(m_1,\dots,m_K)$, where $m_a$ is the number of $a$-colored
 stubs of the vertex. We will assume here that all its moments,
 $\E{m_a}\equiv\sum_{\m} p_{\m} m_a$, $\E{m_am_b}$, etc., are
 defined;
\item a symmetric, non-negative $K\times K$ {\em color preference
 matrix} $\T$, controling the relative abundance, $\sim
 \E{m_a}T_{ab}\E{m_b}$, of edges between different color pairs
 $a,b$. It must satisfy
\beq
\label{T_constr}
	\sum_{b=1}^K T_{ab} \E{m_b} = 1.
\eeq
\end{itemize}
Note that the total degree of a vertex is simply the sum of its
colored degree components; the usual degree distribution is thus also
fixed, and amounts to $p_m=\sum_{\m}\delta(m,\sum_a m_a) p_{\m}$.

%%%%%%%%%%%%%%%% Finite N trunc. %%%%%%%%%%%%%%%%%%
\heading{Truncation to Finite $N$}
We want to implement such an asymptotic model with a specific
$N$.
This can be done e.g. by transforming the CDD into a definite {\em
colored degree sequence}, as described by the number of vertices
$N_{\m} \approx N p_{\m}$ with colored degree $\m$, subject to obvious
constraints such as $m < N$, $\sum_{\m} N_{\m}=N$, and $\sum m N_{\m}$
is even.
Similarly, the matrix $\T$ is used to determine the number of edges
with color-pair $ab$ as $n_{ab} \approx N \E{m_a} T_{ab}
\E{m_b}$. Note that each $ab$-edge is counted twice, as $ab$ and as
$ba$, so the diagonal elements, $n_{aa}$, must be even.
The number of edge endpoints ({\em butts}) with color $a$ becomes $n_a
= \sum_b n_{ab} \approx N \E{m_a} \sum_b T_{ab} \E{m_b}$, and care
must be taken that this matches the corresponding number of stubs,
$\sum_{\m} m_a N_{\m} \approx N \E{m_a}$ -- thus the constraint
(\ref{T_constr}) on $\T$.

This yields a pool of vertices with definite colored degrees and a
pool of edges with definite color pairs, all to be considered
distinguishable.  The set of distinct ways to combine these into a
simple graph with color-matching between butts and stubs defines a set
of colored graphs. By drawing a random member from this set and
neglecting the coloring, the desired truncated CDRG ensemble results.

%%%%%%%%%%%%%%%% Practical implementation: matched pairing and multigraphs %%%%%%%%%%%%%%%%%%
\heading{Implementation in Practice}
When it comes to the practical task of generating random graphs from
this ensemble, the tricky step is that of picking a random member from
the set of colored graphs consistent with definite $N_{\m}$ and
$n_{ab}$. A random stub-pairing method for DRG \cite{BenCan} can be
extended to the case of colored stubs as follows.
\begin{enumerate}
\item \label{pair} For each color $a$, make a complete random
 assignment between the $n_a$ butts of color $a$ and the $n_a$ matching
 stubs, to determine which butt should attach to which stub.
\item While the resulting graph is not simple, repeat step \ref{pair}
\end{enumerate}
Alternatively, the implementation could be done in a fully stochastic
manner, where an extra initial step is to draw $N$ colored degrees
independently from $p_{\m}$, and a pool of edges from $q_{ab} =
\E{m_a}T_{ab}\E{m_b}/\sum_c\E{m_c}$, subject to matching counts of
stubs and butts of each color. In the thermodynamic limit, the result
would be equivalent.
Such a method would be more in line with the identification of CDRG
with the Feynman graphs of zero-dimensional multi-component field
theories, in analogy to the the relation between DRG models and
zero-dimensional scalar field theories \cite{BuCoKr01}

Of course, either generation method is feasible only if the
probability of obtaining a simple graph in each pairing attempt is not
too small. This probability is asymptotically calculable.

%%%%%%%%%%%%%%%% Est of pairing efficiency: alpha, beta %%%%%%%%%%%%%%%%%%
\heading{Pairing Efficiency}
A completely random pairing without the restriction that the resulting
graph be simple yields an ensemble of {\em multigraphs}, i.e. possibly
non-simple graphs where {\em loops} (cycles of length 1) and/or {\em
multiple edges} are allowed.  The efficiency of the above method
depends on the probability to obtain a simple graph, which in turn
depends on the abundance of loops and multiple edges. In a sparse
graph, the probability for an edge between a given pair of nodes
scales as $1/N$, so we expect a finite number both of double edges (a
factor of $N^2$ for the choice of a node-pair, and $1/N^2$ for two
edges), and of loops ($N$ for the choice of node, and $1/N$ for the
edge making a loop).

In fact, we can compute the asymptotically expected number of loops
and double edges in a random pairing to leading order:
\\{\em Loops}: For a single vertex with colored degree $\m$, the
probability that two of its stubs will be connected is given by
$\sum_{ab} (m_a m_b - m_a \delta_{ab}) T_{ab}/2N$. Averaging over $\m$
and summing over the node choice yields the expected number of loops
as $\alpha = \sum_{ab} M_{ab} T_{ab}/2$, i.e.
\beq
\label{alpha}
	\alpha = \frac{1}{2}\tr(\T\M),
\eeq
where $\M = \{M_{ab}\}$ stands for the matrix of moments $\E{m_a m_b -
m_a \delta_{ab}}$.
\\{\em Double edges} Similarly, for an arbitrary pair of nodes with
colored degrees $\m,\m'$, the probability of a double edge
asymptotically amounts to $\sum_{abcd} (m_a m_b - m_a \delta_{ab})
(m'_c m'_d - m'_c \delta_{cd})T_{ac} T_{bd} / (2N^2)$. Averaging over
$\m,\m'$ and summing over the choice of node pair yields the expected
number of double edges as $\beta = \sum_{abcd} M_{ab} M_{cd} T_{ac}
T_{bd} / 4$, i.e.  \beq
\label{beta}
	\beta = \frac{1}{4} \tr (\T\M)^2,
\eeq
while triple edges etc. can be neglected altogether.

In a similar way, the asymptotically expected number of more general
small subgraphs can be computed, which in particular enables the
computation of the expectation of higher powers of the loop and double
edge counts, resulting in the two counts asymptotically behaving as
independent Poissonian random variables.  Hence, the probability of
obtaining a simple graph in the random pairing can be estimated as
\beq
	\text{Prob}(\text{\em simple}) \approx \exp(-\alpha-\beta).
\eeq
As a result, an average of $\sim\exp(\alpha+\beta)$ pairing attempts
is needed, rendering the method feasible for reasonably small
$\alpha+\beta$; in other cases an alternative generation method will
have to be employed, such as starting from an arbitrary colored graph
consistent with $N_{\m},n_{ab}$ and applying a colored extension of a
degree-preserving random rewiring algorithm suggested for DRG
\cite{MasSne02}.

%%%%%%%%%%%%%%%% Conn. comp. analysis %%%%%%%%%%%%%%%%%%
\heading{Connected Component Statistics}
The size-distribution of the connected components (clusters) of a
random graph can be probed by choosing an initial vertex at random and
recursively following edges to new neighbors \cite{MoRe95}. The
sparsity of edges forces a finite set of revealed vertices to form a
{\em tree} in the thermodynamic limit, since cross-linking is
suppressed by factors of $1/N$. Hence, loops and double edges can be
neglected to leading order, and the random color-matched pairing
between stubs and butts reduces to a {\em random branching process}
(branched polymer) based on the rules: {(\bf i)} an edge emanating
from a stub of color $a$ ends in a stub of color $b$ with probability
$T_{ab} \E{m_b}$; {\bf (ii)} given the color $b$ of a stub, it belongs
to a vertex with colored degree $\m$ with probability $m_b
p_{\m}/\E{m_b}$.

The asymptotic random branching process is conveniently described in
terms of a generating function $g(z) = \sum_n P_n z^n$ for the
probability $P_n$ that the connected component being revealed consists
of $n$ vertices. $g(z)$ can be expressed in terms of the corresponding
generating functions $\h(z) = \{h_a(z)\}$ for the number of nodes in a
branch starting from a stub of color $a$.  $g(z)$ and $\h(z)$ satisfy
the recursive relations
\begin{subequations} \label{gh}
\beqa \label{g}
	g(z) &=& z \sum_{\m} p_{\m} \prod_a h_a(z)^{m_a} \equiv z H(\h(z))
\\ \nonumber
	h_a(z) &=& z \sum_b T_{ab} \sum_{\m} p_{\m} m_b \prod_c h_c(z)^{m_c-\delta_{cb}}
\\ \label{h}
	&\equiv & z \sum_b T_{ab} \partial_b H(\h(z)),
\eeqa
\end{subequations}
where $H(\x) = \sum_{\m} p_{\m} \x^\m \equiv \sum_{\m} p_{\m} \prod_a
x_a^{m_a}$ is the multivariate generating function for the CDD, while
$\partial_b$ stands for the derivative with respect to the $b$th
argument of $H$. Eqs. (\ref{gh}) can be derived as follows.
{\em (\ref{g})}: The explicit factor of $z$ accounts for the initial
vertex, while the remainder consists in an average over the colored
degree $\m$ of the initial vertex, of a factor $h_a(z)$ for each stub
of color $a$, accounting for the contribution of the branch starting
in that stub.
{\em (\ref{h})}: Starting from a stub of color $a$, the asymptotic
probability that the other end of the attached edge has color $b$ and
is connected to a vertex having colored degree $\m$ is given by
$T_{ab} p_{\m} m_b$; include a factor $z$ for that vertex, and a
factor $h_c(z)$ for each branch reached via one of its remaining
$(m_c-\delta_{cb})$ stubs of color $c$.

%%%%%%%%%%%%%%%% Percolation threshold & the giant %%%%%%%%%%%%%%%%%%
\heading{Percolation Threshold}
Of particular interest is the value of $g$ for $z=1$: naively we
expect $g(1)=h_a(1)=1$, expressing the normalization of
probability. Indeed, this defines a fixed point of the recurrences
(\ref{gh}), which however may be unstable. The stability can be
analyzed by linearization of eq. (\ref{h}) around $\h(1)=\1$, yielding
the Jacobian matrix $\J$ defined by
\beq
	J_{ab} = \sum_c T_{ac} \left. \partial_c \partial_b H(\h) \right|_{\h=\1},
\eeq
which can be written as $\J=\T\M$ (c.f. eqs. (\ref{alpha},\ref{beta})).

The point is that if an eigenvalue of $\J$ exceeds 1, the naive fixed
point $\h(1)=\1$ turns unstable, signalling {\em supercriticality} of
the branching process. In such a case another fixed point will appear,
and take over as a stable solution with $h_a(1)<1$ yielding $g(1)<1$.
Analogous phenomenona occur in the classic model as well as in IRG and
DRG; the associated probability deficit $1-g(1)$ is interpreted as the
probability of hitting a {\em giant component} asymptotically
containing a finite fraction $1-g(1)$ of the vertices. This
corresponds to a {\em percolating phase}; the percolation threshold is
defined by the largest eigenvalue of $\T\M$ being precisely 1.

%%%%%%%%%%%%%%%% DRG, IRG, RG subsets %%%%%%%%%%%%%%%%%%
\heading{Inclusion of other models}
With a single color, $K=1$, CDRG trivially reduces to DRG, where a
model is based on a given degree distribution $\{p_m\}$, while the
preference matrix $\T$ reduces to a number, which by virtue of the
constraint (\ref{T_constr}) must equal $\E{m}^{-1}$. Equations (\ref{gh})
reduce to the corresponding DRG equations,
\begin{subequations}
\beqa
	g(z) &=& z H(h(z)),
\\
	h(z) &=& z \frac{H'(h(z))}{H'(1)},
\eeqa
\end{subequations}
with $H(x)\equiv \sum_m p_m x^m$ generating $p_m$. The percolating
phase is defined by $J\equiv \E{m(m-1)}/\E{m}>1$, yielding the
well-known $\E{m(m-2)}>0$ \cite{MoRe95}.
%\BS{correct ref!}

The relation to IRG is less trivial: Assume the CDD to be in the form
of a multi-Poissonian mix, i.e.
$
H(\x) = \sum_i r_i \exp\(\sum_a C_{ia} (x_a-1)\)
$.
Define
\begin{subequations}
\label{gihaTc}
\beqa
\label{giha}
	g_i(z) & \equiv & z \exp\(\sum_a C_{ia} (h_a(z)-1)\),
\\
\label{T2c}
	c_{ij} & \equiv & \sum_{ab} C_{ia} T_{ab} C_{jb},
\eeqa
\end{subequations}
in terms of which equations (\ref{gh}) reduce to
\begin{subequations}
\label{ggigigi}
\beqa
\label{ggi}
	g(z) &=& \sum_i r_i g_i(z),
\\
\label{gigi}
	g_i(z) &=& z \exp\( \sum_j r_j c_{ij} \(g_j(z) - 1\) \).
\eeqa
\end{subequations}
Eqs. (\ref{ggigigi}) exactly reproduce the result for $g(z)$ in
an IRG model with $r_i$ taken as the probability of vertex color $i$
and $c_{ij}/N$ the probability of an edge between a pair of vertices
with colors $i,j$ \cite{Sod02}.

Conversely, given an IRG model in terms of $\{r_i,c_{ij}\}$, one can
always find $\{C_{ia},T_{ab}\}$ satisfying eq. (\ref{T2c}) such that
$\sum_a C_{ia} = \sum_j c_{ij} r_j$.

It follows that CDRG contains also ensembles resulting from dynamical
models such as Randomly grown graphs \cite{Call01} and Dynamical
random graphs with memory \cite{Turo02}, that can be described in IRG
\cite{Sod02}, albeit at the cost of infinitely many colors.

%%%%%%%%%%%%%%%% Concluding remarks %%%%%%%%%%%%%%%%%%
\heading{Concluding Remarks}
The above analysis shows that DRG and IRG can be unified into a more
general class of random graph models, defined in terms of a hidden
coloring of stubs and butts, with specified distributions of
color-extended vertex degrees as well of edge colorpairs.  The purpose
of the hidden coloring is to enable a nontrivial correlation structure
in the resulting graphs.

This approach yields a general formalism for a large class of
analytically tractable models on a given degree distribution, where
local and global properties of the resulting graphs are calculable in
the thermodynamic limit. Such a formalism also defines a suitable
target for statistical model inference based on observed structural
properties.

We have here assumed all moments of the degree distribution to exist,
excluding e.g. power behavior. The approach will be extended also to
models with ``fat tails''. These are sensitive to the precise
truncation method and will be treated elsewhere.

A more detailed investigation, addressing aspects and properties of
CDRG models not treated in this letter, is in progress and will the
subject of a forthcoming article, as will the extension to directed
graphs and to degree distributions with power tails.

%%%%%%%%%%%%%%%% Acknowledgment %%%%%%%%%%%%%%%%%%
%\heading{Acknowledgment}
\acknowledgments{This work was in part supported by the Swedish
Foundation for Strategic Research.}

%%%%%%%%%%%%%%%% Refs: %%%%%%%%%%%%%%%%%%
% \bibliography{RG}
% inserted explicitly from bbl-file:
%

\end{document}